%% file: bogoloc_revtex.tex
\documentclass[twocolumn,aps,pra,superscriptaddress,nofootinbib]{revtex4-1}

\usepackage[english]{babel}
\usepackage[utf8]{inputenc}
\usepackage{amsfonts,amsmath,amssymb,bm}
\usepackage{tikz}

\usepackage[colorlinks]{hyperref}

\def\col{} 

\newcommand{\ltr}{l_\text{tr}} 
\newcommand{\lbs}{l_\text{bs}}
 
\newcommand{\vc}[1]{\bm{#1}}

\renewcommand{\k}{{\vc k}}

\newcommand{\p}{{\vc p}}

\newcommand{\w}{{\vc w}}
\newcommand{\x}{{\vc x}}
\newcommand{\xp}{{\vc x'}}
\newcommand{\epn}[1]{\epsilon^0_{#1}}		
\newcommand{\ep}[1]{\epsilon_{#1}}
\DeclareMathSymbol{\Phi}{\mathalpha}{letters}{"08}	

\newcommand{\lloc }{l_{\rm loc}}

\newcommand{\davg}[1]{\overline{#1}}			

\usepackage{enumitem}
\setitemize{topsep=2pt, itemsep=0pt, parsep=2pt, leftmargin=1.5em}
\setenumerate{topsep=2pt, itemsep=0pt, parsep=2pt, leftmargin=1.5em}

\keywords{Quantum transport, Anderson localization, Bose-Einstein condensation, Bogoliubov excitations.}

\begin{document}
\title{Anderson localization of Bogoliubov excitations on quasi-1D strips}
	
\author{Christopher Gaul}
\affiliation{Max-Planck-Institut für Physik Komplexer Systeme, N\"othnitzer Str.~38, 01187 Dresden, Germany}
\author{Pierre Lugan}
\affiliation{Institute of Theoretical Physics, Ecole Polytechnique F\'ed\'erale de Lausanne EPFL, CH-1015 Lausanne, Switzerland}
\author{Cord A. M\"uller}
\affiliation{Institut Non Lin\'eaire de Nice, CNRS and Universit\'e Nice-Sophia Antipolis, 06560 Valbonne, France}
\affiliation{Fachbereich Physik, Universit\"at Konstanz, 78457 Konstanz, Germany}
\begin{abstract}
  Anderson localization of Bogoliubov excitations is studied for
  disordered lattice Bose gases in planar
  quasi--one-dimensional geometries. The inverse localization length
  is computed as function of energy by a numerical transfer-matrix
  scheme, for strips of different widths. These results are described
  accurately by analytical formulas based on a weak-disorder expansion
  of backscattering mean free paths.   
\end{abstract}
\maketitle

\input{bogoloc_v4.tex}

\bibliographystyle{apsrev4-1}
\bibliography{bogoloc}
\end{document}

%% file: bogoloc_v4.tex
\section{Setting, Objectives, and Scope}

Single-particle Anderson localization in quasi-1D geometries 
with correlated disorder is already a challenging problem 
\cite{beenakker_random-matrix_1997,Tessieri2006,Izrailev2012,Herrera-Gonzalez2014}.
With interactions, the situation becomes even more interesting. Here, we study the localization properties of the Bogoliubov quasi-particles (BQP) of weakly disordered Bose gases
at zero temperature on quasi-1D lattices. 
While interactions tend to screen the disorder and thus stabilize extended (quasi-)condensates \cite{petrov_regimes_2000} 
for weak disorder \cite{Lee1990,Mueller2012}, BQPs are expected to be localized irrespective of their energy or disorder strength in low dimensions~\cite{john_localization_1983, gurarie_bosonic_2003, lugan_anderson_2007}, thus emulating noninteracting particles in the orthogonal Wigner-Dyson universality class~\cite{evers_anderson_2008}. Yet, BQPs differ qualitatively from noninteracting particles because of their collective, phonon-like behavior at low energy. Moreover, they experience a randomness mediated by the inhomogeneous 
condensate
background, which responds nonlinearly and nonlocally to the bare disorder~\cite{sanchez-palencia_smoothing_2006, Gaul2011_bogoliubov_long, Lugan2011}. This interplay has been extensively examined in 1D~\cite{gurarie_excitations_2008, fontanesi_mean-field_2010, lellouch_localization_2014, vermersch_anderson_2014}, where direct backscattering provides the only pathway to localization.

\begin{figure}
 \includegraphics{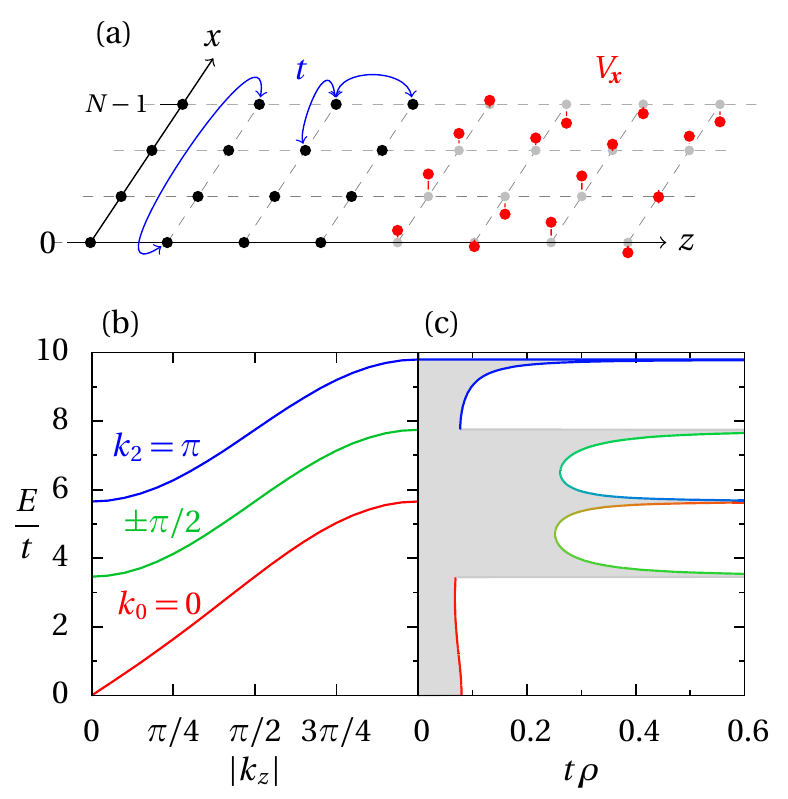}
 \caption{\label{cleanfig}\col 
  System configuration for the $4 \times 1$ geometry.  
  (a) Sketch of the discrete strip geometry, with hopping amplitudes indicated on the left and on-site disorder on the right. 
  (b) Clean lattice Bogoliubov dispersion showing contributions from the transverse modes $k_j$, $(j=0,\dots,3)$.  
  (c) Clean excitation DOS [cf.\ Eq.\ \eqref{freeDOS}] with 1D divergencies at channel openings. The color code indicates the relative contribution of the individual transverse modes. 
  All distances are given in units of lattice spacing. Panels (b) and (c) are drawn for $U n/t=2.0$.
 }
\end{figure}

In this Letter, we discuss the localization length of BQPs on quasi-1D planar lattices of transverse width $N\times 1$ (shown in Fig.~\ref{cleanfig}a). While the techniques developed below also apply to $N_x\times N_y$ bars, strips provide the simplest realization of a multi-channel geometry, where phase-coherent scattering between modes is the rule. 
The bosons are described by the Bose-Hubbard (BH) model with on-site
interaction $U>0$, hopping~$t$ (with periodic boundary conditions across the strip) and on-site disorder~$V_\x$. The random potential~$V_\x$ is drawn from a Gaussian distribution and has no spatial correlation on the lattice scale, $\davg{V_\x V_{\x'}} = \delta_{\x \x'}V^2$,
where the bar denotes disorder averaging and $V^2$ is the on-site variance. 
For each realization of disorder, the mean-field (MF) ground-state density
$n_\x$ is determined by the condensate amplitude 
$\Phi_\x=\sqrt{n_\x}$ that solves the discrete Gross-Pitaevskii (GP) equation
\begin{align}\label{eq:GP}
-t\sum_{\langle \x \xp \rangle}\Phi_{\xp}+(2t+V_{\x})\Phi_{\x}+U\Phi_{\x}^3 &= \mu \Phi_{\x},
\end{align}
where the sum runs over the nearest neighbors of $\x$. 
The chemical potential $\mu$ controls the average occupation $n=\overline{n_\x}$. In the limit of large occupations~\cite{mora_extension_2003}, an expansion of the BH Hamiltonian around the MF solution yields the Bogoliubov-de Gennes (BdG) equations
\begin{align}\label{eq:BdG}
-t\sum_{\langle \x \xp \rangle}u_{\xp}+[2t+V_{\x}+2U n_{\x}-\mu]u_{\x}+U n_{\x} v_{\x}&=E u_{\x}\nonumber\\
-t\sum_{\langle \x \xp \rangle}v_{\xp}+[2t+V_{\x}+2U n_{\x}-\mu]v_{\x}+U n_{\x} u_{\x}&=-E v_{\x},
\end{align}
where $u_{\x}$ and $v_{\x}$ are the BQP particle and hole components at energy~$E$. These Bogoliubov excitations determine ground-state properties as well as the 
\mbox{(thermo-)}\! \!\,\mbox{dynamical} response of the system.

In the remainder of this paper, we compute the dependence of the BQP localization length on the excitation energy. We first present a numerical transfer-matrix method that provides us with precise estimates of the localization length over a wide range of parameters (section~\ref{Sec:num}). Then we develop an analytical framework that combines random-matrix theory~\cite{beenakker_random-matrix_1997} with a microscopic transport theory (section~\ref{Sec:analytics}). In particular, we find that the predictions using a lattice Boltzmann mean-free path are in excellent agreement with the numerics.

\section{Numerical transfer-matrix calculations}
\label{Sec:num}

Our numerical calculations of the localization length follow a standard transfer-matrix scheme~\cite{pichard_finite-size_1981, kramer_localization:_1993, beenakker_random-matrix_1997, slevin_critical_2014}: 
The BdG equations (\ref{eq:BdG}) are cast into the recursive form $\w_{i+1}=M_i\w_i$, where the $4N$-component vector $\w_i$ encodes the values of $u_\x$ and $v_\x$ on two neighboring lattice slices at $z=i$ and $z=i-1$ (throughout the article, distances are expressed in units of lattice spacing). The transfer matrix $M_i$ is parametrized by $E$ as well as the values of $V_\x$ and $n_{\x}$ on slice $i$. The propagation of the BdG excitations along~$z$, from an arbitrary initial condition~$\w_0$, then simply amounts to matrix-vector multiplication. Via this procedure, the estimator $\ln \|\w_L\|/L$ yields, with probability one, the largest Lyapunov exponent $\lambda_1$ in the limit $L\to\infty$. The $r$ largest Lyapunov exponents $\lambda_1>\dots>\lambda_r$ are retrieved similarly, by propagating a frame of $r$ linearly independent initial vectors $\w_0^{[j]}$ ($j=1,\dots,r$), enforcing their orthogonality during propagation, and monitoring their exponential growth (to maintain numerical accuracy, we performed a Gram-Schmidt orthonormalization every 8 sites). As the BQP Lyapunov exponents come in pairs of opposite sign, $2N$ vectors are required to calculate the smallest positive Lyapunov exponent $\lambda=\lambda_{2N}$, identified as the inverse localization length $1/\lloc$~\cite{beenakker_random-matrix_1997, slevin_critical_2014}.

One crucial advantage of the transfer-matrix approach is the self-averaging of the $\ln \|\w_L^{[j]}\|/L$ estimators, whose relative fluctuations decrease as $1/\sqrt{L}$. The propagation of \textit{one} set of initial vectors over a large distance $L$ thus suffices to compute the Lyapunov spectrum with the desired precision.
Such a scheme is easy to implement with an uncorrelated random potential, as the purely local random values of $V_\x$ can be generated on the fly. In the case of BQPs, however, the GP equation~(\ref{eq:GP}) implies a global minimization problem that is already intractable for the size required to achieve only 10\% accuracy on the Lyapunov exponents. 
In our calculations, we chained quasi-1D segments of length $L_0=2^{14}$ and, on each segment, solved Eq.~(\ref{eq:GP}) at fixed density $n$ with a conjugate-gradient technique~\cite{modugno_bose-einstein_2003, saliba_superfluid-insulator_2014}. 
While each segment interface introduces a slight mismatch in $\mu$, and also in $n_\x$ over typically $\sqrt{t/(U n)}$ sites, the impact of these artifacts on $\lambda$ vanishes with increasing $L_0$. We checked that all the results presented here are converged with respect to the choice of $L_0$. To ensure a sufficient scrambling of initial data and speed up estimations, we chained up to 64 segments and averaged the $\ln\|\w_L^{[j]}\|$ data over 200 independent propagations.

\begin{figure}
	\includegraphics[width=\linewidth]{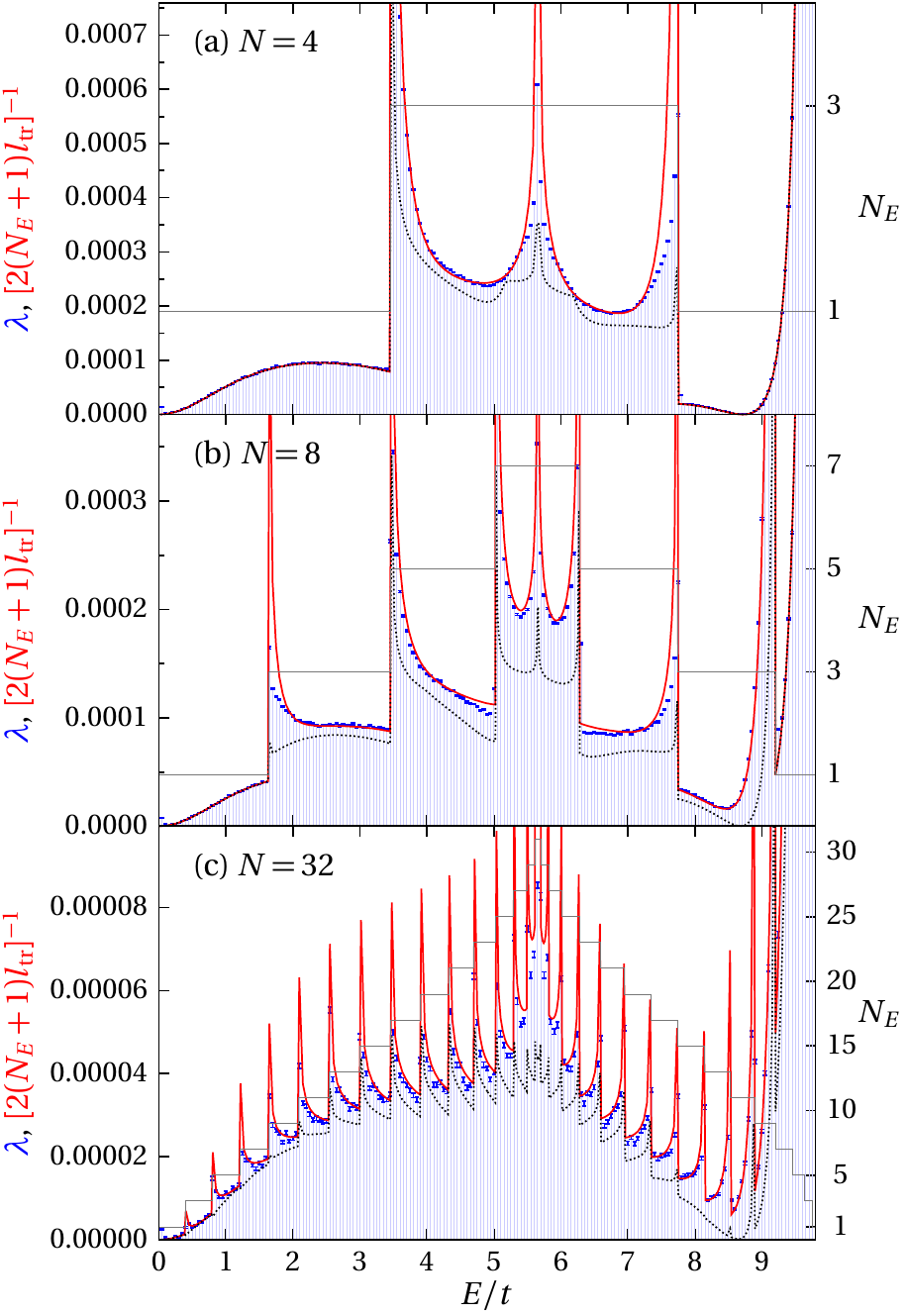}
 \caption{\label{figLocLengths}\col 
 Inverse localization length $\lambda$ (error bars indicate the error of the mean after averaging over 500 configurations in (a) and (b) and 225 in (c), and are hardly appreciable on the scale of the plot) and its estimates from transport theory as a function of energy. 
 The dashed line is {$[(N_E+1)\lbs]^{-1}$} from Eq.~\eqref{lbs},
 and the solid red line is $[2(N_E+1)\ltr]^{-1}$ from Eq.~\eqref{ltr}. The staircase-like line indicates the number of open channels $N_E$ (scale on the right border).
 Parameters: hopping $U n = 2 t \approx\mu$, i.e., the transition from sound waves to particle excitations is around $E/t=2$. Disorder strength $V=0.1 U n$.}
\end{figure}

Figure~\ref{figLocLengths} shows the inverse localization length $\lambda$ as function of energy for various strip widths $N$. For all $N$ we observe two important features. First, $\lambda$ diverges at the inner band edges, where the channels of the clean system open or close (cf.~Fig.~\ref{cleanfig}c) and thus contribute a divergent 1D density of states (DOS), which results in strong scattering and thus localization on short length scales. 
Second, for an equal number of open channels and once inner divergences are disregarded, $\lambda$ displays a similar trend as in 1D~\cite{lugan_anderson_2007}, increasing with $E$ in the phonon regime $E\lesssim U n$, and decreasing in the free-particle regime $E\gtrsim U n$. 

\section{Analytical transport theory}\label{sec:analytical}
\label{Sec:analytics}

For quasi-1D strips, random-matrix theory \cite{beenakker_random-matrix_1997} predicts that the
localization length $\lloc = 2(N_E+1) \ltr$ is proportional to the
transport mean free path $\ltr$. 
Here, $N_E$ is the number of open channels at energy $E$. In this section, we validate this
prediction using two different analytical expressions for the mean
free path. 

The clean system is discrete translation invariant, and therefore the
free-particle 
dispersion relation is diagonal in momentum, $\epn{\k} = 2 t \sum_{i=x,z}  [1 - \cos(k_i) ]$. The dispersion of Bogoliubov excitations then
reads as usual  
$\ep{\k} = [\epn{\k}(\epn{\k} + 2 U n)]^{1/2}$. On the strip, the momentum $\k = (k_j,k_z)$ is transverse
quantized as $k_j= 2\pi j/N $ under periodic boundary conditions with
$j=0,\dots,N-1$, thus defining the $N$ possible channels. 
In terms of the longitudinal group velocity $v_\k = \partial
\ep{\k}/\partial k_z$ (all velocities here and below refer to the $z$-component), the DOS per unit length and transverse size is expressed as  
\begin{equation}
\rho(E) = \frac{1}{N} \sum_{j=0}^{N-1} \int_{-\pi}^\pi \frac{d k_z}{2\pi} \delta (E-\ep{\k}) =
\frac{1}{2\pi N} \sum_{\k \in S_E} |v_\k|^{-1}. 
\label{freeDOS}
\end{equation} 
$S_E$ denotes the energy shell in momentum space, namely the set of points $\k$ such that $\ep{\k}=E$. Its cardinality $|S_E|=2N_E$ counts both forward and backward propagating channels with $k_z>0$ and $k_z<0$, respectively. 
The clean DOS is plotted in Fig.~\ref{cleanfig}c. 
The opening and closing of each channel produces a characteristic van Hove divergence. Near the divergence, essentially only one channel matters. In between these singularities several channels mix, as indicated by the color coding in the plot. 
 

The disorder potential produces elastic scattering out of a given mode $\k$ with a rate 
$\gamma_\k = 2\pi \sum_\p \delta(\epsilon_\k - \ep{\p}) \davg{|W_{\k \p}|^2} $.
Here $W_{\k \p} = V_{\k-\p} w_{\k \p}$ is a (first-order) matrix element of the
disordered Bogoliubov Hamiltonian, 
namely the bare disorder Fourier component $V_{\k-\p}$ 
dressed by an envelope 
\begin{equation}
w_{\k \p}
{=} \frac{(\epsilon^0_\k + U n) \epsilon^0_{\k - \p} - 2U n \epsilon^0_\k }{\epsilon_\k (2 U n + \epsilon^0_{\k - \p}) } \qquad (\epn{\k}=\epn{\p}) 
\label{onshellEnvelope}
\end{equation} 
that accounts for the underlying nonlinearities \cite{Gaul2011_bogoliubov_long,Gaul2013_bogolattice}.
[Eq.\ \eqref{onshellEnvelope} is the on-shell value $ w_{\k \p}^{(1)}|_{\epn{\k}=\epn{\p}}$ given by Eqs.\ (13) -- (15) in \cite[]{Gaul2013_bogolattice} or, equivalently, the lattice version of Eq.~(14) in \cite[]{Gaul2008}.]
The associated elastic mean free path is then the group velocity
divided by the scattering rate. 
Since localization 
along the $z$ direction can only be caused by scattering events 
that flip the momentum $z$-component, we consider the
\emph{backscattering} mean free path
 \begin{equation}
\lbs(E)^{-1}  
= \min_{\k\in S_E} \frac{V^2}{N |v_\k| } {\sum_{\p}}'  |v_\p|^{-1} w_{\k \p}^2 \ . \label{lbs}
\end{equation}
The primed sum runs only over those momenta $\p\in S_{\ep{\k}}$ with $z$-components of opposite sign,
$\mathrm{sgn}(p_z k_z)=-1$. 
Moreover, we take the shortest inverse mean free path 
at a given energy $E$ since the numerics produce the smallest localization exponent. The resulting prediction $[(N_E+1)\lbs (E)]^{-1}$ is plotted 
as the dashed line in Fig.~\ref{figLocLengths}. 
There is already some overall qualitative agreement, which becomes even quantitative at low and high energies, where
only a single channel contributes. And indeed, in this  case ($N_E=1$) one has 
$2\pi N \rho(E) = 2|v_k|^{-1}$ at $\ep{k}=E$, and thus the estimate for the Lyapunov exponent $
 (2\lbs)^{-1} = V^2w_{k(-k)}^2/(2 N v_k^2)
$ agrees with the established theoretical result \cite{lugan_anderson_2007,Lugan2011,Gaul2011_bogoliubov_long}.%
\footnote{In the 1D case with $\pm k$ along $z$ and quadratic dispersion $\epn{2k} = 4\epn{k}$, one has $w_{k(-k)} = {\epn{k}}/{\ep{k}}$ 
and thus $[2 \lbs(\ep{k})]^{-1} = V^2 S_k^2/[2 N (v_k^0)^2 ]$
in terms of 
the free-particle velocity $v_k^0 = \partial \epn{k}/\partial k$
and 
$S_k=\epn{k}/(U n+\epn{k})$,
which agrees with 
\cite[Eq.~(11)]{lugan_anderson_2007}.}

The number $N_E$ of open channels is also plotted in each figure. With more open channels, the
estimate involving $\lbs$ becomes less reliable, which becomes especially obvious near the center of
plots b) and c) for the $8\times 1$ and $32\times 1$ geometries.  
The agreement deteriorates even further if only
diagonal backscattering into the same channel ($p_j=k_j$ and $p_z=-k_z$) is taken into account. 

We therefore try to improve on our estimate and
consider the lattice version of the full Boltzmann transport mean free
path, which can be derived within linear-response quantum
transport theory \cite{Kuhn2007,KuhnPhd2007}. With the notations
introduced above, this gives 
\begin{align}
\label{ltr}
\ltr(E)^{-1}
	&= \frac{V^2}{2\pi N^2 \rho(E) \langle v^2 \rangle_{E}^{3/2}} 
	\sum_{ 
		{\k,\p \in S_E}}\!
		\frac{		v_\p  (v_\p - v_\k) 
			  w_{\k\p}^2
                       }{|v_\p| |v_\k|}
\end{align}
Here, 
$ \langle v^2 \rangle_{E} = \sum_{\k \in S_E} |v_\k|/
 [{2\pi}N \rho(E)]$ denotes an energy-shell average. The
usual factor $(1 -
\cos \theta)$ in scattering angle $\theta$ under the momentum integral
(see, e.g., \cite[Fig.\
6]{Gaul2011_bogoliubov_long} and \cite[Eq.\ (6)]{Kuhn2005})
is now represented by $(v_\p-v_\k)$. The factors of
$|v_\k|$ and $|v_\p|$ in the denominator stem from the
$k_z$-integrals over on-shell spectral functions, analogous to the
second equality in \eqref{freeDOS}. 

For a strict 1D problem, i.e., for
a single open channel ($N_E=1$), this expression 
reduces to $ 2\ltr  
= \lbs$, as it should \cite[Eq.~(152)]{beenakker_random-matrix_1997}.
But also when several channels are open, 
the agreement with the numerical data is clearly excellent, 
as shown in Fig.\ \ref{figLocLengths}. 
Remarkably, Eq.~\eqref{ltr} performs especially well in between the 1D resonances, 
where the coherent coupling between channels has to be described rather accurately. 

Of course, the result \eqref{ltr} is perturbative in nature, 
essentially valid for weak enough disorder ($V \ll U n$), and thus 
only holds if the (quasi-)condensate is extended. 
For much stronger disorder, the condensate fragments and the Bose fluid enters 
the non-superfluid, Bose-glass phase~\cite{fontanesi_fragmentation_2011, saliba_superfluid-insulator_2014}.

\section{Summary and Outlook}
\label{Sec:concl}

In conclusion, we have calculated the localization length
for Bogoliubov excitations of Bose-Einstein condensates on
disordered lattices with a planar quasi-1D geometry. Numerical data
from a transfer-matrix computation are very well reproduced by an
analytical formula adapted from continuum transport theory. Our approach could extend to larger systems $N\times 1$  (and $N\times N$) with $N \to \infty$  using finite-size scaling and thus yield the full 2D (and 3D) localization lengths, with the aim to track down possible mobility edges for BQPs in dimensions higher than 1.

%% file: bogoloc_revtex.bbl
\begin{thebibliography}{30}%
\makeatletter
\providecommand \@ifxundefined [1]{%
 \@ifx{#1\undefined}
}%
\providecommand \@ifnum [1]{%
 \ifnum #1\expandafter \@firstoftwo
 \else \expandafter \@secondoftwo
 \fi
}%
\providecommand \@ifx [1]{%
 \ifx #1\expandafter \@firstoftwo
 \else \expandafter \@secondoftwo
 \fi
}%
\providecommand \natexlab [1]{#1}%
\providecommand \enquote  [1]{``#1''}%
\providecommand \bibnamefont  [1]{#1}%
\providecommand \bibfnamefont [1]{#1}%
\providecommand \citenamefont [1]{#1}%
\providecommand \href@noop [0]{\@secondoftwo}%
\providecommand \href [0]{\begingroup \@sanitize@url \@href}%
\providecommand \@href[1]{\@@startlink{#1}\@@href}%
\providecommand \@@href[1]{\endgroup#1\@@endlink}%
\providecommand \@sanitize@url [0]{\catcode `\\12\catcode `\$12\catcode
  `\&12\catcode `\#12\catcode `\^12\catcode `\_12\catcode `\%12\relax}%
\providecommand \@@startlink[1]{}%
\providecommand \@@endlink[0]{}%
\providecommand \url  [0]{\begingroup\@sanitize@url \@url }%
\providecommand \@url [1]{\endgroup\@href {#1}{\urlprefix }}%
\providecommand \urlprefix  [0]{URL }%
\providecommand \Eprint [0]{\href }%
\providecommand \doibase [0]{http://dx.doi.org/}%
\providecommand \selectlanguage [0]{\@gobble}%
\providecommand \bibinfo  [0]{\@secondoftwo}%
\providecommand \bibfield  [0]{\@secondoftwo}%
\providecommand \translation [1]{[#1]}%
\providecommand \BibitemOpen [0]{}%
\providecommand \bibitemStop [0]{}%
\providecommand \bibitemNoStop [0]{.\EOS\space}%
\providecommand \EOS [0]{\spacefactor3000\relax}%
\providecommand \BibitemShut  [1]{\csname bibitem#1\endcsname}%
\let\auto@bib@innerbib\@empty
\bibitem [{\citenamefont {Beenakker}(1997)}]{beenakker_random-matrix_1997}%
  \BibitemOpen
  \bibfield  {author} {\bibinfo {author} {\bibfnamefont {C.~W.~J.}\
  \bibnamefont {Beenakker}},\ }\href {\doibase 10.1103/RevModPhys.69.731}
  {\bibfield  {journal} {\bibinfo  {journal} {Rev. Mod. Phys.}\ }\textbf
  {\bibinfo {volume} {69}},\ \bibinfo {pages} {731} (\bibinfo {year}
  {1997})}\BibitemShut {NoStop}%
\bibitem [{\citenamefont {Tessieri}\ and\ \citenamefont
  {Izrailev}(2006)}]{Tessieri2006}%
  \BibitemOpen
  \bibfield  {author} {\bibinfo {author} {\bibfnamefont {L.}~\bibnamefont
  {Tessieri}}\ and\ \bibinfo {author} {\bibfnamefont {F.~M.}\ \bibnamefont
  {Izrailev}},\ }\href {http://stacks.iop.org/0305-4470/39/11717} {\bibfield
  {journal} {\bibinfo  {journal} {J. Phys. A: Math. Gen.}\ }\textbf {\bibinfo
  {volume} {39}},\ \bibinfo {pages} {11717} (\bibinfo {year}
  {2006})}\BibitemShut {NoStop}%
\bibitem [{\citenamefont {Izrailev}\ \emph {et~al.}(2012)\citenamefont
  {Izrailev}, \citenamefont {Krokhin},\ and\ \citenamefont
  {Makarov}}]{Izrailev2012}%
  \BibitemOpen
  \bibfield  {author} {\bibinfo {author} {\bibfnamefont {F.}~\bibnamefont
  {Izrailev}}, \bibinfo {author} {\bibfnamefont {A.}~\bibnamefont {Krokhin}}, \
  and\ \bibinfo {author} {\bibfnamefont {N.}~\bibnamefont {Makarov}},\ }\href
  {\doibase 10.1016/j.physrep.2011.11.002} {\bibfield  {journal} {\bibinfo
  {journal} {Phys. Rep.}\ }\textbf {\bibinfo {volume} {512}},\ \bibinfo {pages}
  {125 } (\bibinfo {year} {2012})}\BibitemShut {NoStop}%
\bibitem [{\citenamefont {Herrera-Gonz\'alez}\ \emph
  {et~al.}(2014)\citenamefont {Herrera-Gonz\'alez}, \citenamefont
  {M\'endez-Berm\'udez},\ and\ \citenamefont
  {Izrailev}}]{Herrera-Gonzalez2014}%
  \BibitemOpen
  \bibfield  {author} {\bibinfo {author} {\bibfnamefont {I.~F.}\ \bibnamefont
  {Herrera-Gonz\'alez}}, \bibinfo {author} {\bibfnamefont {J.~A.}\ \bibnamefont
  {M\'endez-Berm\'udez}}, \ and\ \bibinfo {author} {\bibfnamefont {F.~M.}\
  \bibnamefont {Izrailev}},\ }\href {\doibase 10.1103/PhysRevE.90.042115}
  {\bibfield  {journal} {\bibinfo  {journal} {Phys. Rev. E}\ }\textbf {\bibinfo
  {volume} {90}},\ \bibinfo {pages} {042115} (\bibinfo {year}
  {2014})}\BibitemShut {NoStop}%
\bibitem [{\citenamefont {Petrov}\ \emph {et~al.}(2000)\citenamefont {Petrov},
  \citenamefont {Shlyapnikov},\ and\ \citenamefont
  {Walraven}}]{petrov_regimes_2000}%
  \BibitemOpen
  \bibfield  {author} {\bibinfo {author} {\bibfnamefont {D.~S.}\ \bibnamefont
  {Petrov}}, \bibinfo {author} {\bibfnamefont {G.~V.}\ \bibnamefont
  {Shlyapnikov}}, \ and\ \bibinfo {author} {\bibfnamefont {J.~T.~M.}\
  \bibnamefont {Walraven}},\ }\href {\doibase 10.1103/PhysRevLett.85.3745}
  {\bibfield  {journal} {\bibinfo  {journal} {Phys. Rev. Lett.}\ }\textbf
  {\bibinfo {volume} {85}},\ \bibinfo {pages} {3745} (\bibinfo {year}
  {2000})}\BibitemShut {NoStop}%
\bibitem [{\citenamefont {Lee}\ and\ \citenamefont {Gunn}(1990)}]{Lee1990}%
  \BibitemOpen
  \bibfield  {author} {\bibinfo {author} {\bibfnamefont {D.~K.~K.}\
  \bibnamefont {Lee}}\ and\ \bibinfo {author} {\bibfnamefont {J.~M.~F.}\
  \bibnamefont {Gunn}},\ }\href {\doibase 10.1088/0953-8984/2/38/004}
  {\bibfield  {journal} {\bibinfo  {journal} {J. Phys.: Condens. Matter}\
  }\textbf {\bibinfo {volume} {2}},\ \bibinfo {pages} {7753} (\bibinfo {year}
  {1990})}\BibitemShut {NoStop}%
\bibitem [{\citenamefont {M\"uller}\ and\ \citenamefont
  {Gaul}(2012)}]{Mueller2012}%
  \BibitemOpen
  \bibfield  {author} {\bibinfo {author} {\bibfnamefont {C.~A.}\ \bibnamefont
  {M\"uller}}\ and\ \bibinfo {author} {\bibfnamefont {C.}~\bibnamefont
  {Gaul}},\ }\href {http://stacks.iop.org/1367-2630/14/i=7/a=075025} {\bibfield
   {journal} {\bibinfo  {journal} {New J. Phys.}\ }\textbf {\bibinfo {volume}
  {14}},\ \bibinfo {pages} {075025} (\bibinfo {year} {2012})}\BibitemShut
  {NoStop}%
\bibitem [{\citenamefont {John}\ \emph {et~al.}(1983)\citenamefont {John},
  \citenamefont {Sompolinsky},\ and\ \citenamefont
  {Stephen}}]{john_localization_1983}%
  \BibitemOpen
  \bibfield  {author} {\bibinfo {author} {\bibfnamefont {S.}~\bibnamefont
  {John}}, \bibinfo {author} {\bibfnamefont {H.}~\bibnamefont {Sompolinsky}}, \
  and\ \bibinfo {author} {\bibfnamefont {M.~J.}\ \bibnamefont {Stephen}},\
  }\href {\doibase 10.1103/PhysRevB.27.5592} {\bibfield  {journal} {\bibinfo
  {journal} {Phys. Rev. B}\ }\textbf {\bibinfo {volume} {27}},\ \bibinfo
  {pages} {5592} (\bibinfo {year} {1983})}\BibitemShut {NoStop}%
\bibitem [{\citenamefont {Gurarie}\ and\ \citenamefont
  {Chalker}(2003)}]{gurarie_bosonic_2003}%
  \BibitemOpen
  \bibfield  {author} {\bibinfo {author} {\bibfnamefont {V.}~\bibnamefont
  {Gurarie}}\ and\ \bibinfo {author} {\bibfnamefont {J.~T.}\ \bibnamefont
  {Chalker}},\ }\href {\doibase 10.1103/PhysRevB.68.134207} {\bibfield
  {journal} {\bibinfo  {journal} {Phys. Rev. B}\ }\textbf {\bibinfo {volume}
  {68}},\ \bibinfo {pages} {134207} (\bibinfo {year} {2003})}\BibitemShut
  {NoStop}%
\bibitem [{\citenamefont {Lugan}\ \emph {et~al.}(2007)\citenamefont {Lugan},
  \citenamefont {Clément}, \citenamefont {Bouyer}, \citenamefont {Aspect},\
  and\ \citenamefont {Sanchez-Palencia}}]{lugan_anderson_2007}%
  \BibitemOpen
  \bibfield  {author} {\bibinfo {author} {\bibfnamefont {P.}~\bibnamefont
  {Lugan}}, \bibinfo {author} {\bibfnamefont {D.}~\bibnamefont {Clément}},
  \bibinfo {author} {\bibfnamefont {P.}~\bibnamefont {Bouyer}}, \bibinfo
  {author} {\bibfnamefont {A.}~\bibnamefont {Aspect}}, \ and\ \bibinfo {author}
  {\bibfnamefont {L.}~\bibnamefont {Sanchez-Palencia}},\ }\href {\doibase
  10.1103/PhysRevLett.99.180402} {\bibfield  {journal} {\bibinfo  {journal}
  {Phys. Rev. Lett.}\ }\textbf {\bibinfo {volume} {99}},\ \bibinfo {pages}
  {180402} (\bibinfo {year} {2007})}\BibitemShut {NoStop}%
\bibitem [{\citenamefont {Evers}\ and\ \citenamefont
  {Mirlin}(2008)}]{evers_anderson_2008}%
  \BibitemOpen
  \bibfield  {author} {\bibinfo {author} {\bibfnamefont {F.}~\bibnamefont
  {Evers}}\ and\ \bibinfo {author} {\bibfnamefont {A.~D.}\ \bibnamefont
  {Mirlin}},\ }\href {\doibase 10.1103/RevModPhys.80.1355} {\bibfield
  {journal} {\bibinfo  {journal} {Rev. Mod. Phys.}\ }\textbf {\bibinfo {volume}
  {80}},\ \bibinfo {pages} {1355} (\bibinfo {year} {2008})}\BibitemShut
  {NoStop}%
\bibitem [{\citenamefont
  {Sanchez-Palencia}(2006)}]{sanchez-palencia_smoothing_2006}%
  \BibitemOpen
  \bibfield  {author} {\bibinfo {author} {\bibfnamefont {L.}~\bibnamefont
  {Sanchez-Palencia}},\ }\href@noop {} {\bibfield  {journal} {\bibinfo
  {journal} {Phys. Rev. A}\ }\textbf {\bibinfo {volume} {74}},\ \bibinfo
  {pages} {053625} (\bibinfo {year} {2006})}\BibitemShut {NoStop}%
\bibitem [{\citenamefont {Gaul}\ and\ \citenamefont
  {M{\"u}ller}(2011)}]{Gaul2011_bogoliubov_long}%
  \BibitemOpen
  \bibfield  {author} {\bibinfo {author} {\bibfnamefont {C.}~\bibnamefont
  {Gaul}}\ and\ \bibinfo {author} {\bibfnamefont {C.~A.}\ \bibnamefont
  {M{\"u}ller}},\ }\href {\doibase 10.1103/PhysRevA.83.063629} {\bibfield
  {journal} {\bibinfo  {journal} {Phys. Rev. A}\ }\textbf {\bibinfo {volume}
  {83}},\ \bibinfo {pages} {063629} (\bibinfo {year} {2011})}\BibitemShut
  {NoStop}%
\bibitem [{\citenamefont {Lugan}\ and\ \citenamefont
  {Sanchez-Palencia}(2011)}]{Lugan2011}%
  \BibitemOpen
  \bibfield  {author} {\bibinfo {author} {\bibfnamefont {P.}~\bibnamefont
  {Lugan}}\ and\ \bibinfo {author} {\bibfnamefont {L.}~\bibnamefont
  {Sanchez-Palencia}},\ }\href {\doibase 10.1103/PhysRevA.84.013612} {\bibfield
   {journal} {\bibinfo  {journal} {Phys. Rev. A}\ }\textbf {\bibinfo {volume}
  {84}},\ \bibinfo {pages} {013612} (\bibinfo {year} {2011})}\BibitemShut
  {NoStop}%
\bibitem [{\citenamefont {Gurarie}\ \emph {et~al.}(2008)\citenamefont
  {Gurarie}, \citenamefont {Refael},\ and\ \citenamefont
  {Chalker}}]{gurarie_excitations_2008}%
  \BibitemOpen
  \bibfield  {author} {\bibinfo {author} {\bibfnamefont {V.}~\bibnamefont
  {Gurarie}}, \bibinfo {author} {\bibfnamefont {G.}~\bibnamefont {Refael}}, \
  and\ \bibinfo {author} {\bibfnamefont {J.~T.}\ \bibnamefont {Chalker}},\
  }\href {\doibase 10.1103/PhysRevLett.101.170407} {\bibfield  {journal}
  {\bibinfo  {journal} {Phys. Rev. Lett.}\ }\textbf {\bibinfo {volume} {101}},\
  \bibinfo {pages} {170407} (\bibinfo {year} {2008})}\BibitemShut {NoStop}%
\bibitem [{\citenamefont {Fontanesi}\ \emph {et~al.}(2010)\citenamefont
  {Fontanesi}, \citenamefont {Wouters},\ and\ \citenamefont
  {Savona}}]{fontanesi_mean-field_2010}%
  \BibitemOpen
  \bibfield  {author} {\bibinfo {author} {\bibfnamefont {L.}~\bibnamefont
  {Fontanesi}}, \bibinfo {author} {\bibfnamefont {M.}~\bibnamefont {Wouters}},
  \ and\ \bibinfo {author} {\bibfnamefont {V.}~\bibnamefont {Savona}},\ }\href
  {\doibase 10.1103/PhysRevA.81.053603} {\bibfield  {journal} {\bibinfo
  {journal} {Phys. Rev. A}\ }\textbf {\bibinfo {volume} {81}},\ \bibinfo
  {pages} {053603} (\bibinfo {year} {2010})}\BibitemShut {NoStop}%
\bibitem [{\citenamefont {Lellouch}\ and\ \citenamefont
  {Sanchez-Palencia}(2014)}]{lellouch_localization_2014}%
  \BibitemOpen
  \bibfield  {author} {\bibinfo {author} {\bibfnamefont {S.}~\bibnamefont
  {Lellouch}}\ and\ \bibinfo {author} {\bibfnamefont {L.}~\bibnamefont
  {Sanchez-Palencia}},\ }\href {\doibase 10.1103/PhysRevA.90.061602} {\bibfield
   {journal} {\bibinfo  {journal} {Phys. Rev. A}\ }\textbf {\bibinfo {volume}
  {90}},\ \bibinfo {pages} {061602} (\bibinfo {year} {2014})}\BibitemShut
  {NoStop}%
\bibitem [{\citenamefont {Vermersch}\ \emph {et~al.}(2014)\citenamefont
  {Vermersch}, \citenamefont {Delande},\ and\ \citenamefont
  {Garreau}}]{vermersch_anderson_2014}%
  \BibitemOpen
  \bibfield  {author} {\bibinfo {author} {\bibfnamefont {B.}~\bibnamefont
  {Vermersch}}, \bibinfo {author} {\bibfnamefont {D.}~\bibnamefont {Delande}},
  \ and\ \bibinfo {author} {\bibfnamefont {J.~C.}\ \bibnamefont {Garreau}},\
  }\href {http://arxiv.org/abs/1410.2587} {\bibfield  {journal} {\bibinfo
  {journal} {{arXiv}:1410.2587}\ } (\bibinfo {year} {2014})}\BibitemShut
  {NoStop}%
\bibitem [{\citenamefont {Mora}\ and\ \citenamefont
  {Castin}(2003)}]{mora_extension_2003}%
  \BibitemOpen
  \bibfield  {author} {\bibinfo {author} {\bibfnamefont {C.}~\bibnamefont
  {Mora}}\ and\ \bibinfo {author} {\bibfnamefont {Y.}~\bibnamefont {Castin}},\
  }\href {\doibase 10.1103/PhysRevA.67.053615} {\bibfield  {journal} {\bibinfo
  {journal} {Phys. Rev. A}\ }\textbf {\bibinfo {volume} {67}},\ \bibinfo
  {pages} {053615} (\bibinfo {year} {2003})}\BibitemShut {NoStop}%
\bibitem [{\citenamefont {Pichard}\ and\ \citenamefont
  {Sarma}(1981)}]{pichard_finite-size_1981}%
  \BibitemOpen
  \bibfield  {author} {\bibinfo {author} {\bibfnamefont {J.~L.}\ \bibnamefont
  {Pichard}}\ and\ \bibinfo {author} {\bibfnamefont {G.}~\bibnamefont
  {Sarma}},\ }\href {\doibase 10.1088/0022-3719/14/21/004} {\bibfield
  {journal} {\bibinfo  {journal} {J. Phys. C}\ }\textbf {\bibinfo {volume}
  {14}},\ \bibinfo {pages} {L617} (\bibinfo {year} {1981})}\BibitemShut
  {NoStop}%
\bibitem [{\citenamefont {Kramer}\ and\ \citenamefont
  {MacKinnon}(1993)}]{kramer_localization:_1993}%
  \BibitemOpen
  \bibfield  {author} {\bibinfo {author} {\bibfnamefont {B.}~\bibnamefont
  {Kramer}}\ and\ \bibinfo {author} {\bibfnamefont {A.}~\bibnamefont
  {MacKinnon}},\ }\href {\doibase 10.1088/0034-4885/56/12/001} {\bibfield
  {journal} {\bibinfo  {journal} {Rep. Prog. Phys.}\ }\textbf {\bibinfo
  {volume} {56}},\ \bibinfo {pages} {1469} (\bibinfo {year}
  {1993})}\BibitemShut {NoStop}%
\bibitem [{\citenamefont {Slevin}\ and\ \citenamefont
  {Ohtsuki}(2014)}]{slevin_critical_2014}%
  \BibitemOpen
  \bibfield  {author} {\bibinfo {author} {\bibfnamefont {K.}~\bibnamefont
  {Slevin}}\ and\ \bibinfo {author} {\bibfnamefont {T.}~\bibnamefont
  {Ohtsuki}},\ }\href {\doibase 10.1088/1367-2630/16/1/015012} {\bibfield
  {journal} {\bibinfo  {journal} {New J. Phys.}\ }\textbf {\bibinfo {volume}
  {16}},\ \bibinfo {pages} {015012} (\bibinfo {year} {2014})}\BibitemShut
  {NoStop}%
\bibitem [{\citenamefont {Modugno}\ \emph {et~al.}(2003)\citenamefont
  {Modugno}, \citenamefont {Pricoupenko},\ and\ \citenamefont
  {Castin}}]{modugno_bose-einstein_2003}%
  \BibitemOpen
  \bibfield  {author} {\bibinfo {author} {\bibfnamefont {M.}~\bibnamefont
  {Modugno}}, \bibinfo {author} {\bibfnamefont {L.}~\bibnamefont
  {Pricoupenko}}, \ and\ \bibinfo {author} {\bibfnamefont {Y.}~\bibnamefont
  {Castin}},\ }\href {\doibase 10.1140/epjd/e2003-00015-y} {\bibfield
  {journal} {\bibinfo  {journal} {Eur. Phys. J. D}\ }\textbf {\bibinfo {volume}
  {22}},\ \bibinfo {pages} {235} (\bibinfo {year} {2003})}\BibitemShut
  {NoStop}%
\bibitem [{\citenamefont {Saliba}\ \emph {et~al.}(2014)\citenamefont {Saliba},
  \citenamefont {Lugan},\ and\ \citenamefont
  {Savona}}]{saliba_superfluid-insulator_2014}%
  \BibitemOpen
  \bibfield  {author} {\bibinfo {author} {\bibfnamefont {J.}~\bibnamefont
  {Saliba}}, \bibinfo {author} {\bibfnamefont {P.}~\bibnamefont {Lugan}}, \
  and\ \bibinfo {author} {\bibfnamefont {V.}~\bibnamefont {Savona}},\ }\href
  {\doibase 10.1103/PhysRevA.90.031603} {\bibfield  {journal} {\bibinfo
  {journal} {Phys. Rev. A}\ }\textbf {\bibinfo {volume} {90}},\ \bibinfo
  {pages} {031603} (\bibinfo {year} {2014})}\BibitemShut {NoStop}%
\bibitem [{\citenamefont {{Gaul}}\ and\ \citenamefont
  {{M{\"u}ller}}(2013)}]{Gaul2013_bogolattice}%
  \BibitemOpen
  \bibfield  {author} {\bibinfo {author} {\bibfnamefont {C.}~\bibnamefont
  {{Gaul}}}\ and\ \bibinfo {author} {\bibfnamefont {C.~A.}\ \bibnamefont
  {{M{\"u}ller}}},\ }\href {\doibase 10.1140/epjst/e2013-01755-9} {\bibfield
  {journal} {\bibinfo  {journal} {Eur. Phys. J. ST}\ }\textbf {\bibinfo
  {volume} {217}},\ \bibinfo {pages} {69} (\bibinfo {year} {2013})}\BibitemShut
  {NoStop}%
\bibitem [{\citenamefont {Gaul}\ and\ \citenamefont
  {M{\"u}ller}(2008)}]{Gaul2008}%
  \BibitemOpen
  \bibfield  {author} {\bibinfo {author} {\bibfnamefont {C.}~\bibnamefont
  {Gaul}}\ and\ \bibinfo {author} {\bibfnamefont {C.~A.}\ \bibnamefont
  {M{\"u}ller}},\ }\href {\doibase 10.1209/0295-5075/83/10006} {\bibfield
  {journal} {\bibinfo  {journal} {Europhys. Lett.}\ }\textbf {\bibinfo {volume}
  {83}},\ \bibinfo {pages} {10006} (\bibinfo {year} {2008})}\BibitemShut
  {NoStop}%
\bibitem [{\citenamefont {Kuhn}\ \emph {et~al.}(2007)\citenamefont {Kuhn},
  \citenamefont {Sigwarth}, \citenamefont {Miniatura}, \citenamefont
  {Delande},\ and\ \citenamefont {M{\"u}ller}}]{Kuhn2007}%
  \BibitemOpen
  \bibfield  {author} {\bibinfo {author} {\bibfnamefont {R.}~\bibnamefont
  {Kuhn}}, \bibinfo {author} {\bibfnamefont {O.}~\bibnamefont {Sigwarth}},
  \bibinfo {author} {\bibfnamefont {C.}~\bibnamefont {Miniatura}}, \bibinfo
  {author} {\bibfnamefont {D.}~\bibnamefont {Delande}}, \ and\ \bibinfo
  {author} {\bibfnamefont {C.~A.}\ \bibnamefont {M{\"u}ller}},\ }\href
  {http://www.iop.org/EJ/article/1367-2630/9/6/161/njp7_6_161.pdf} {\bibfield
  {journal} {\bibinfo  {journal} {New J. Phys.}\ }\textbf {\bibinfo {volume}
  {9}},\ \bibinfo {pages} {161} (\bibinfo {year} {2007})}\BibitemShut {NoStop}%
\bibitem [{\citenamefont {Kuhn}(2007)}]{KuhnPhd2007}%
  \BibitemOpen
  \bibfield  {author} {\bibinfo {author} {\bibfnamefont {R.}~\bibnamefont
  {Kuhn}},\ }\emph {\bibinfo {title} {Coherent Transport of Matter Waves in
  Disordered Optical Potentials}},\ \href
  {http://opus.ub.uni-bayreuth.de/volltexte/2007/287/} {Ph.D. thesis},\
  \bibinfo  {school} {Universit{\"a}t Bayreuth \& Universit{\'e} de Nice
  Sophia--Antipolis} (\bibinfo {year} {2007})\BibitemShut {NoStop}%
\bibitem [{\citenamefont {Kuhn}\ \emph {et~al.}(2005)\citenamefont {Kuhn},
  \citenamefont {Miniatura}, \citenamefont {Delande}, \citenamefont
  {Sigwarth},\ and\ \citenamefont {M{\"u}ller}}]{Kuhn2005}%
  \BibitemOpen
  \bibfield  {author} {\bibinfo {author} {\bibfnamefont {R.~C.}\ \bibnamefont
  {Kuhn}}, \bibinfo {author} {\bibfnamefont {C.}~\bibnamefont {Miniatura}},
  \bibinfo {author} {\bibfnamefont {D.}~\bibnamefont {Delande}}, \bibinfo
  {author} {\bibfnamefont {O.}~\bibnamefont {Sigwarth}}, \ and\ \bibinfo
  {author} {\bibfnamefont {C.~A.}\ \bibnamefont {M{\"u}ller}},\ }\href
  {\doibase 10.1103/PhysRevLett.95.250403} {\bibfield  {journal} {\bibinfo
  {journal} {Phys. Rev. Lett.}\ }\textbf {\bibinfo {volume} {95}},\ \bibinfo
  {pages} {250403} (\bibinfo {year} {2005})}\BibitemShut {NoStop}%
\bibitem [{\citenamefont {Fontanesi}\ \emph {et~al.}(2011)\citenamefont
  {Fontanesi}, \citenamefont {Wouters},\ and\ \citenamefont
  {Savona}}]{fontanesi_fragmentation_2011}%
  \BibitemOpen
  \bibfield  {author} {\bibinfo {author} {\bibfnamefont {L.}~\bibnamefont
  {Fontanesi}}, \bibinfo {author} {\bibfnamefont {M.}~\bibnamefont {Wouters}},
  \ and\ \bibinfo {author} {\bibfnamefont {V.}~\bibnamefont {Savona}},\ }\href
  {\doibase 10.1103/PhysRevA.83.033626} {\bibfield  {journal} {\bibinfo
  {journal} {Phys. Rev. A}\ }\textbf {\bibinfo {volume} {83}},\ \bibinfo
  {pages} {033626} (\bibinfo {year} {2011})}\BibitemShut {NoStop}%
\end{thebibliography}%
